%                   G. 't Hooft  macros version 2000
\newread\testifexists
\def\GetIfExists #1 {\immediate\openin\testifexists=#1
    \ifeof\testifexists\immediate\closein\testifexists\else
    \immediate\closein\testifexists\input #1\fi}
\def\epsffile#1{Figure: #1}     %%%

\immediate\openin\testifexists=e
    \ifeof\testifexists\immediate\closein\testifexists\else
    \immediate\closein\testifexists\input e\fipsf

\magnification= \magstep1  % or use \magstephalf
\tolerance=1600
\parskip=5pt
\baselineskip= 5 true mm \mathsurround=1pt
\font\smallrm=cmr8  
\font\medrm=cmr9 \font\medit=cmti9 
\font\bigbf=cmbx12
    \def\Bbb#1{\setbox0=\hbox{$\tt #1$}  \copy0\kern-\wd0\kern .1em\copy0}
    \immediate\openin\testifexists=a
    \ifeof\testifexists\immediate\closein\testifexists\else
    \immediate\closein\testifexists\input a\fimssym.def %% for \Bbb A - Z %%

\def\secbreak{\vskip12pt plus .6in \penalty-200\vskip -2pt plus -.4in}
 %\prefbreak{distance}
\def\ref#1{${\,}^{\hbox{\smallrm #1}}$}
   \def\newsect#1{\secbreak\noindent{\bf #1}\medskip}
   \def\br{\vfil\break}
\def\hugeskip{\vskip12mm plus 3mm}
\def\Narrower{\par\narrower\noindent}   % never again use TeX's awful \narrower
\def\Endnarrower{\par\leftskip=0pt \rightskip=0pt}
\def\br{\hfil\break}    \def\ra{\rightarrow}

\def\cl{\centerline}    
\def\ni{\noindent}      \def\pa{\partial}   \def\dd{{\rm d}}
\def\tl{\tilde}            \def\ket{\rangle}

\def\a{\alpha}      \def\b{\beta}         
\def\d{\delta}      \def\D{\Delta}  \def\e{\varepsilon}
               \def\L{\Lambda}
\def\m{\mu}         \def\f{\phi}            \def\vv{\varphi}
\def\n{\nu}             
     \def\s{\sigma}  
\def\t{\tau}        \def\th{\theta}  
              
\def\w{\omega}

\def\fn#1{\ifcase\noteno\def\fnchr{*}\or\def\fnchr{\dagger}\or\def
    \fnchr{\ddagger}\or\def\fnchr{\medrm\S}\or\def\fnchr{\|}\or\def
    \fnchr{\medrm\P}\fi\footnote{$^{\fnchr}$}
    {\scrunch#1\toe}\ifnum\noteno>4\global\advance\noteno by-6\fi
    \global\advance\noteno by 1}
    \def\scrunch{\baselineskip=11 pt \medrm}
    \def\toe{\vphantom{$p_\big($}}
    \newcount\noteno
    %    footnote with alternating symbol   %

\def\fract#1#2{{\textstyle{#1\over#2}}}
\def\ffract#1#2{{\raise .35 em\hbox{$\scriptstyle#1$}\kern-.25em/
    \kern-.2em\lower .22 em\hbox{$\scriptstyle#2$}}}

\def\half{\fract12}  \def\quart{\fract14}

\def\bbf#1{\setbox0=\hbox{$#1$} \kern-.025em\copy0\kern-\wd0
    \kern.05em\copy0\kern-\wd0 \kern-.025em\raise.0433em\box0}
    % boldface in math mode.

\def\deff{\ {\buildrel{\rm def}\over{=}}\ }
\def\eq{\ =\ }
%{\nopagenumbers %
{\ }\vglue 1truecm
\rightline{SPIN-2001/11}\rightline{ITP-UU-01/18}
\rightline{hep-th/0105105} \hugeskip \cl{\bigbf QUANTUM MECHANICS
AND DETERMINISM\fn{Presented at PASCOS 2001, Eighth International
Symposium on Particles, Strings and Cosmology, University of North
Carolina at Chapel Hill, April 10-15, 2001.}} \hugeskip

\cl{Gerard 't Hooft }
\bigskip
\cl{Institute for Theoretical Physics} \cl{Utrecht University,
Leuvenlaan 4} \cl{ 3584 CC Utrecht, the Netherlands}
\smallskip
\cl{and}
\smallskip
\cl{Spinoza Institute} \cl{Postbox 80.195} \cl{3508 TD Utrecht,
the Netherlands}
\smallskip\cl{e-mail: \tt g.thooft@phys.uu.nl}
\cl{internet: \tt http://www.phys.uu.nl/\~{}thooft/} \hugeskip
\ni{\bf Abstract}\Narrower It is shown how to map the quantum
states of a system of free scalar particles one-to-one onto the
states of a completely deterministic model. It is a classical
field theory with a large (global) gauge group. The mapping is now
also applied to free Maxwell fields. Lorentz invariance is
demonstrated. \Endnarrower \hugeskip
\newsect{1. Introduction.} Numerous attempts to reconcile General
Relativity with Quantum Mechanics appear to lead to descriptions
of space and time at the Planck scale where notions such as
locality, unitarity and causality are in jeopardy. In particular
Super String theories only allow for computations of on-shell
amplitudes, so that the local nature of the laws of physics
becomes obscure. A further complication is the holographic
principle\fn{The general notion underlying this principle can be
found in Ref\ref1. The word ``holography" was mentioned in print,
I think, first in Ref\ref2.}, which states that the total
dimensionality of Hilbert space is controlled rather by the
surface than by the volume of a given region in space.

As this situation is fundamentally different from what we have
both in non-relativistic Quantum Mechanics and in relativistic
quantum field theories, including the Standard Model, it appears
to be quite reasonable to reconsider the foundations of Quantum
Mechanics itself. Exactly to what extent a completely
deterministic hidden variable theory is feasible is still not
understood; leaving this question aside for the time being, it
may nevertheless be instructive to formulate some general and
useful mathematical starting points.

In a previous paper\ref{3} it was shown how a quantum field theory
of free scalar particles can be mapped onto a classical field
theory. The ontological states of the classical theory form the
basis of a Hilbert space, and these states evolve in accordance
with a Schr\"odinger equation that coincides with the
Schr\"odinger equation of the quantum theory\ref4. Thus, the
classical model can be used to `interpret' the quantum mechanics
of the quantum theory. The price one pays is two-fold: first, a
large invariance group (`gauge group') must be introduced in the
classical system, and we must restrict ourselves to those
observables which are invariant under the transformations of this
group. Since the gauge transformations are non-local, the
observables are not obviously well-defined locally. Secondly, the
procedure is known to work only in some very special cases: either
massless non-interacting fermions (in $\leq 4$ space-time
dimensions), or free scalar particles (massless or massive, in any
number of space-time dimensions).

There are reasons to suspect however that more general models
might exist that allow such mappings, and it may not be entirely
unreasonable to conjecture that physically reasonable models will
eventually be included. In this contribution, we show how to
handle free Maxwell fields, a result that is not quite trivial
because of our desire to keep rotational covariance.

Secondly, we show how to handle Lorentz transformations. The
examples treated here turn out to be Lorentz-invariant classical
theories. For the Maxwell case, this is not a trivial derivation,
requiring an interplay between our ontological gauge group and the
Weyl gauge group.

In Sects.~2 and~3, we briefly resume the argument from Ref\ref3
that the quantum harmonic oscillator corresponds to classical
circular motion, and that free scalar quantized fields can be
linked to a classical theory with a special kind of gauge
invariance, such that all its Fourier modes are purely circular
degrees of freedom.

Lorentz transformations are discussed in Sect.~4. The Maxwell case
is handled in Sect.~5. We add a brief discussion of the situation
for fermions in Sect.~6, and the interpretation of our gauge
transformations in terms of a dissipation theory (Sect.~7).

\newsect{2. Harmonic oscillators.}

We start with a {\it deterministic\/} system, consisting of a set
of $N$ states, $$\{(0),(1),\cdots,(N-1)\}$$ on a circle. Time is
discrete, the unit time steps having length $\t$ (the continuum
limit is left for later). The evolution law is:
    $$t\ra t+\t\quad:\qquad (\n)\ra(\n+1\,{\rm\ mod}\, N)\,.\eqno(2.1)$$
Introducing a basis for a Hilbert space spanned by the states
$(\n)$, the evolution operator can be written as
    $$U(\D t=\t)\ =\ e^{-iH\t}\ =e^{-\textstyle{\pi i\over
    N\vphantom{_g}}}\cdot\pmatrix{0&&&&1\cr 1\,&0\cr &1&0\cr
    &&\ddots&\ddots\cr&&&1&0\cr }\ .\eqno(2.2)$$
The phase factor in front of the matrix is of little importance;
it is there for future convenience. Its eigenstates are denoted as
$|n\ket$, $n=0,\cdots,N-1$.

This law can be represented by a Hamiltonian using the notation
of quantum physics:
    $$H|n\ket={2\pi(n+\half)\over N\t}|n\ket\,.\eqno(2.3)$$
The $\half$ comes from the aforementioned phase factor. Next, we
apply the algebra of the $SU(2)$ generators $L_x$, $L_y$ and
$L_z$, so we write
    $$N\deff 2\ell+1\quad,\qquad n\deff m+\ell\quad,\qquad
    m=-\ell,\cdots,\ell\ .\eqno(2.4)$$
Using the quantum numbers $m$ rather than $n$ to denote the
eigenstates, we have
    $$H|m\ket={2\pi(m+\ell+\half )\over (2\ell+1)\t}|m\ket\qquad\hbox{or}\qquad
    H={\textstyle{2\pi\over (2\ell+1)\t}}\,(L_z+\ell+\half)\ .
    \eqno(2.5)$$
This Hamiltonian resembles the harmonic oscillator Hamiltonian
with angular frequencies \hbox{$\w=2\pi/(2\ell+1)\t$}, except for
the fact that there is an upper bound for the energy. This upper
bound disappears in the continuum limit, if $\ell\ra\infty$,
$\t\downarrow 0$. Using $L_x$ and $L_y$, we can make the
correspondence more explicit. Write
    $$\eqalign{ L_\pm|m\ket&\deff \sqrt{\ell(\ell+1)-m(m\pm1)}\ |m\pm1\ket\
    ;\cr  L_\pm&\deff L_x\pm iL_y \quad;\qquad[L_i,L_j]=i\e_{ijk}L_k\
    ,}\eqno(2.6)$$
and define
    $$\hat x\deff\a L_x\quad,\qquad \hat p\deff\b L_y\quad;\qquad
    \a\deff\sqrt{\t\over\pi}\quad,\qquad\b\deff{-2\over
    2\ell+1}\sqrt{\pi\over\t} \ .\eqno(2.7)$$
The commutation rules are
    $$[\hat x,\hat p]=\a\b iL_z=i(1-{\t\over\pi}H)\,,\eqno(2.8)$$
and since
    $$L_x^2+L_y^2+L_z^2=\ell(\ell+1)\,,\eqno(2.9)$$ we have
    $$H=\half\w^2 \hat x^2+\half \hat p^2+ {\t\over2\pi}\left({\w^2\over 4}+
    H^2\right)\,. \eqno(2.10)$$
The coefficients $\a$ and $\b$ in Eqs.~(2.7) have been tuned to
give (2.8) and (2.10) their most desirable form.

Now consider the continuum limit, $\t\downarrow 0$, with
$\w=2\pi/(2\ell+1)\t$ fixed, for those states for which the energy
stays limited. We see that the commutation rule (2.8) for $\hat x$
and $\hat p$ becomes the conventional one, and the Hamiltonian
becomes that of the conventional harmonic oscillator. There are no
other states than the legal ones, and their energies are bounded,
as can be seen not only from (2.10) but rather from the original
definition (2.5). Note that, in the continuum limit, both $\hat x$
and $\hat p$ become continuous operators.

The way in which these operators act on the `primordial' or
`ontological' states $(\n)$ of Eq.~(2.1) can be derived from
(2.6) and (2.7), if we realize that the states $|m\ket$ are just
the discrete Fourier transforms of the states $(\n)$. This way,
also the relation between the eigenstates of $\hat x$ and $\hat p$
and the states $(\n)$ can be determined. Only in a fairly crude
way, $\hat x$ and $\hat p$ give information on where on the
circle our ontological object is; both $\hat x$ and $\hat p$
narrow down the value of $\n$ of our states $(\n)$.

The  most important conclusion from this section is that there is
a close relationship between the quantum harmonic oscillator and
the classical particle moving along a circle. The period of the
oscillator is equal to the period of the trajectory along the
circle. We started our considerations by having time discrete, and
only a finite number of states. This is because the continuum
limit is a rather delicate one. One cannot directly start with the
continuum because then the Hamiltonian does not seem to be bounded
from below.

The price we pay for a properly bounded Hamiltonian is the square
root in Eq.~(2.6); it may cause complications when we attempt to
introduce interactions, but this is not the subject of this paper.

\newsect{3. Free scalar particles.}
Now consider the Klein-Gordon equation describing a quantized
field $\f$,
    $$(\D-\m^2)\f-\ddot\f=0\,,\eqno(3.2)$$
where the dots refer to time differentiation. It represents
coupled harmonic oscillators. We decouple them by diagonalizing
the equation, that is, we consider the Fourier modes. For each
Fourier mode, with given $\pm\vec k$, there are two quantum
harmonic oscillators, because the Fourier coefficients are
complex.

In principle, the procedure to be followed may appear to be
straightforward: introduce a dynamical degree of freedom moving on
a circle, with angular frequencies $\w(\vec k)=(\vec
k^2+\m^2)^{1/2}$, for each of these modes. The question is, how to
introduce a model for which the Fourier modes contain just such
degrees of freedom; an ordinary classical field, to be denoted as
$\{\vv(\vec x,t)\}$, would contain Fourier modes, $\hat\vv(\vec
k,t)$, which do not only have a circular degree of freedom, but
also an amplitude:
    $$\vv(\vec x,t)\deff (2\pi)^{-3/2}\int\dd^3\vec k\,\hat\vv(\vec
    k,t)\,e^{i\vec k\cdot\vec x}\ .\eqno(3.1)$$
$\hat\vv(\vec k,t)$ are all classical oscillators. They are not
confined to the circle, but the real parts, $\Re(\hat\vv(\vec
k,t))$, and the imaginary parts, $\Im(\hat\vv(\vec k,t))$ of
every Fourier mode each move in a two-dimensional phase space. If
we want to reproduce the quantum system, we have to replace these
two-dimensional phase spaces by one-dimensional circles.

The trick that will be employed is to introduce a new kind of
gauge invariance. In a given classical oscillator
    $$\dot x=y\quad,\qquad \dot y=-\w^2 x\ ,\eqno(3.2)$$
we want to declare that the amplitude is unobservable; only the
{\it phase\/} is physical. So, we introduce transformations of the
form
    $$x\ra \L x\quad,\qquad y\ra \L y\ ,\eqno(3.3)$$
where the transformation parameter $\L$ is a real, positive
number.

Now, however, a certain amount of care is needed. We do not want
to destroy translation invariance. A space translation would mix
up the real and imaginary parts of $\hat\vv(\vec k,t)$ and
$\dot{\hat\vv}(\vec k,t)$. This is why it is not advised to start
from the real part and the imaginary part, and subject these to
the transformations (3.3) separately. Rather, we note that, at
each $\vec k$, there are two oscillatory modes, a positive and a
negative frequency. Thus, in general,
    $$\eqalign{\hat\vv(\vec k,t)&=A(\vec k)\,e^{i\w t}+B(\vec
    k)\,e^{-i\w t}\,;\cr \dot{\hat\vv}(\vec k,t)&=i\w A(\vec
    k)\,e^{i\w t}-i\w B(\vec k)\,e^{-i\w t}\,,}\eqno(3.4)$$
where $\w=(\vec k^2+\m^2)^{1/2}$. It is these amplitudes that we
may subject to the transformations (3.3), so that we only keep the
circular motions $e^{\pm i\w t}$.

Thus, we introduce the `gauge transformation'
    $$\eqalign{A(\vec k)&\ra  R_1(\vec k)A(\vec k)\ ,\cr
    B(\vec k)&\ra R_2(\vec k)B(\vec k)\ ,}\eqno(3.5)$$
where $R_1(\vec k)$ and $R_2(\vec k)$ are {\it real, positive\/}
functions of $\vec k$. The {\it only\/} quantities invariant
under these two transformations are the phases of $A$ and $B$,
which is what we want. In terms of $\vv$ and $\dot\vv$, the
transformation reads:
    $$\eqalign{\vv&\ra \textstyle
    {R_1+R_2\over 2}\,\vv+\textstyle{R_1-R_2\over
    2i\w}\,\dot\vv\ ,\cr \dot\vv&\ra \textstyle{R_1+R_2\over
    2}\,\dot\vv+\textstyle{i\w(R_1-R_2)\over 2}\,\vv\ .}\eqno(3.6)$$

Writing
    $$\eqalign{\hat K_1(\vec k)&=\textstyle{R_1+R_2\over 2}\ ,\cr
    \hat K_2(\vec k)&=\textstyle{R_2-R_1\over 2\w}\ ,}\eqno(3.7)$$
and Fourier transforming, we see that, in coordinate space,
    $$\eqalign{\vv(\vec x,t)&\ra\int\dd^3\vec
    y\,\big( K_1(\vec y)\,\vv(\vec x+\vec y,t)+K_2(\vec
    y)\,\dot\vv(\vec x+\vec y,t)\big)\,,\cr  \dot\vv (\vec
    x,t)&\ra\int\dd^3\vec y\,\big( K_1(\vec y)\,\dot\vv(\vec x+\vec
    y,t)+K_2(\vec y)\,(\D-\m^2)\vv(\vec x+\vec y,t)\big)\,.
    }\eqno(3.8)$$
Since $R_1$ and $R_2$ are real in $\vec k$-space, the kernels
$K_1$ and $K_2$ obey the constraints:
    $$K_1(\vec y)=K_1(-\vec y)\quad;\qquad K_2(\vec y)=-K_2(-\vec
    y)\,,\eqno(3.9)$$
and they are time-independent. The classical Klein-Gordon equation
is obviously invariant under these transformations. The fact that
$R_1$ and $R_2$ are constrained to be positive, amounts to
limiting oneself to the homogeneous part of the gauge
group.\fn{Alternatively, one could consider dropping such
limitations, which would require dividing the angular frequencies
$\w$ by a factor 2, because then the phase angles are defined
{\medit modulo\/} $180^\circ$ only. We believe however that the
resulting theory will be physically unacceptable.}

Note that the requirement that physical observables are invariant
under these transformations, essentially reduces the set of
physically observable dynamical degrees of freedom by half. This
is essentially what Quantum Mechanics does: in Quantum Mechanics,
a complete specification of only coordinates {\it or\/} only
momenta suffices to fix the degrees of freedom at a given time;
in a classical theory, one would normally have to specify
coordinates as well as momenta.

Although the gauge group is a very large one, it should still be
characterized as a {\it global\/} gauge group, since the kernels
$K_i$ depend on $\vec y$ and not on $\vec x$, and they are time
independent.
\newsect{4. Lorentz invariance.}

The gauge transformation (3.8) is non-local in space, but local in
time. Hence, one could question whether the selection of gauge
invariant observables is Lorentz invariant. Of course, rotational
invariance is evident from the notation, even though the kernels
$K_i(\vec y)$ are in general {\it not\/} rotationally invariant.
The fact that Lorentz invariance still holds, is guaranteed by the
equations of motion of the field $\vv(\vec x,t)$. This is derived
as follows.

Consider an infinitesimal Lorentz rotation in the $x$-direction,
given the fields $\vv(\vec x,0)$ and $\dot\vv(\vec x,0)$ at $t=0$:
    $$\vv'=\vv+\e\d\vv\quad,\qquad\dot\vv'=\dot\vv+\e\d\dot\vv\,,\eqno(4.1)$$
where $\e$ is infinitesimal. Of course, $\vv$ transforms as a
Lorentz scalar and $\dot\vv$ as a Lorentz vector. We have
    $$\eqalign{\d\vv\eq x\pa_t\vv+t\pa_x\vv&\eq x\dot\vv+t\pa_x\vv\ ;\cr
    \d\dot\vv\eq x\pa_t\dot\vv+t\pa_x\dot\vv+\pa_x\vv&\eq x(\D-\m^2)\vv+
    t\pa_x\dot\vv+\pa_x\vv\ .}\eqno(4.2)$$
This transforms the gauge transformation (3.8) into
    $$\eqalign{\f'&\ra K'_1\vv'+K'_2\dot\vv'\ ;\cr
    \dot\vv'&\ra K'_1\dot\vv'+K'_2\vv'\ .}\eqno(4.3)$$
with $K'_i=K_i+\d K_i$, and
    $$\eqalign{\d K_1&=(\D-\m^2)(x\,K_2-K_2\,x)-\pa_x K_2\ ;\cr
    \d K_2&=x\,K_1-K_1\,x\ .}\eqno(4.4)$$
In these expressions, $K_i\vv$ stands short for $\int\dd^3\vec
y\,K_i(\vec y)\vv(x+\vec y,t)$, and $x\,K_2-K_2\,x$ replaces
$K_2(\vec y)$ by $-y_x\,K_2(\vec y)$. We easily read off from
these expressions that the transformed kernels $K'_i(\vec y)$ have
the correct symmetry properties (3.9) under reflection in $\vec
y$.

This explicit calculation confirms a more abstract argument, which
simply observes that multiplying the amplitudes $A$ and $B$ in
Eq.~(3.4) by real numbers should be a Lorentz invariant procedure.

We conclude from this section that, if in one Lorentz frame
observables are required to be invariant under all gauge rotations
(3.8) that obey the symmetry constraints (3.9), then this
continues to hold in any other Lorentz frame. The theory is
Lorentz invariant.

\newsect{5. Maxwell fields.}

Handling theories with multiples of bosonic fields may seem to be
straightforward. In general, however, simple symmetries such as
isospin or rotational invariance are not guaranteed. Take the
simple example of a vector field $\vec A$ in 3-space. We now have
three quantum harmonic oscillators for each $\vec k$, which would
have to be linked to three circular degrees of freedom. This would
lead us to a gauge transformation of the form (3.8) for each of
the three vector components $A_i$ separately. Clearly this is not
rotationally invariant. The kernels $K_i(\vec y)$ would have to be
replaced by symmetric matrices instead, but this would give us a
gauge group that is far too large.

In some special cases, this difficulty can be removed, and an
example at hand is the Maxwell theory. We consider pure Maxwell
fields without any charged sources. Again, we first diagonalize
the Hamiltonian by going to Fourier space. As is well-known, the
photon has two polarizations, so at every $\vec k$, there are now
two quantum oscillators with positive $\w$ and two with negative
$\w$.

Let us choose $\vec k$ to lie along the $z$-axis. Then $\hat
A_x(\vec k,t)$ and $\hat A_y(\vec k,t)$ should rotate into one
another under rotations about the $z$-axis. Can we choose two
rotators in such a way that this rotation symmetry is kept?

Consider again the discretized case, where we have a finite time
interval $\t$. The two rotators have the same angular velocity
$\w=\pm|\vec k|$. The Hamiltonian (normalized to $2(\half\w)$ for
the ground state) is therefore
    $$H=|\w|(L_z^{(1)}+L_z^{(2)}+2\ell+1\,.\eqno(5.1)$$
These states are illustrated in Fig.~1, for the case $\ell=3$.
\midinsert\cl{\epsffile{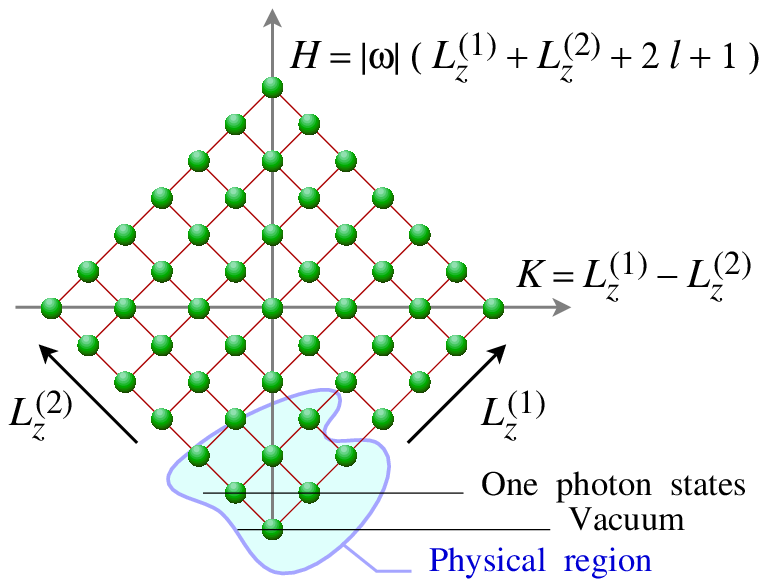}}\Narrower Figure~1.
Two-boson states with a $U(1)$ exchange symmetry. $H$ generates
time translations: $\th_1\ra\th_1+\w\t$, $\th_2\ra\th_2+\w\t$,
whereas $K$ generates rotations about the $\vec k$ axis:
$\th_1\ra\th_1+\t$, $\th_2\ra\th_2-\t$.\Endnarrower\endinsert

States with $H=(n+1)\w$ contain $n$-photons, whose total spin
ranges from $-n$ to $n$. This is the range of the operator
$K=L_z^{(1)}-L_z^{(2)}$. Thus, we have to arrange the states in
accordance with their spin values along the axis of the momentum
variable $\vec k$ (which we take to be the $z$-axis). This way,
the energy eigenstates of our circular degrees of freedom will
automatically also be eigenstates of transverse rotations. We are
lead to consider the field modes $A_\pm$, each of which has
positive and negative frequencies:
    $$A_\pm\deff A_x\pm iA_y\quad;\qquad\hat
    A_\pm(\vec k,t)\ra A_\pm e^{i\w t}+B_\pm e^{-i\w t}\,.\eqno(5.2)$$
We can be assured that this decomposition is covariant under
rotations about the $z$-axis.

As before, we restrict ourselves to the phase components of these
degrees of freedom, while the amplitudes themselves are subject to
gauge transformations. These gauge transformations are
    $$A_\pm\ra K_\pm A_\pm\quad,\qquad B_\pm\ra L_\pm B_\pm\ ,\eqno(5.3)$$
with $K_\pm$ and $L_\pm$ both real and positive.

At $t=0$, we get, plugging (5.2) into (5.3):
    $$\eqalign{\hat A_i &\ra\hat K_1\hat A_i+i\e_{ijk}k_j\,\hat K_2\hat
    A_k +i\hat L_1\dot{\hat A_i}-\e_{ijk}k_j\,\hat L_2\dot{\hat A_k}\ ;\cr
    \dot{\hat A_i} &\ra -i\w^2\hat L_1\hat A_i+\e_{ijk}k_j\,\w^2\hat L_2\hat
    A_k +\hat K_1\dot{\hat A_i}+i\e_{ijk}k_j\,\hat K_2\dot{\hat
    A_k}\ ,}\eqno(5.4)$$  where $\w=|k|$ and
    $$ \eqalign{\hat K_1=\quart(K_++K_++L_++L_-)\quad,&\qquad
    \hat K_2={\textstyle{ 1\over 4\,\w}}(-K_++K_--L_++L_-)\ ,\cr
    \hat L_1={\textstyle{ 1\over 4\,\w}}(-K_+-K_-+L_++L_-)\quad,&\qquad
    \hat L_2={\textstyle{ 1\over 4\,\w^2}}(K_+-K_--L_++L_-)\ .}\eqno(5.5)$$
In coordinate space this transformation reads
    $$\eqalign{A_i(\vec x) \ra\int\dd^3\vec y\Big(&K_1(\vec
    y)A_i(\vec x+\vec y)+K_2(\vec y)\e_{ijk}\pa_jA_k(\vec x+\vec
    y)\cr &+L_1(\vec y)\dot A_i(\vec x+\vec y)+L_2(\vec y)\e_{ijk}\pa_j
    \dot A_k(\vec x+\vec y)\Big)\ ;\cr
    \dot A_i(\vec x) \ra\int\dd^3\vec y\Big(&L_1(\vec y)\D A_i(\vec
    x+\vec y)+L_2(\vec y)\e_{ijk}\pa_j\D A_k(\vec x+\vec y)\cr
    &+K_1(\vec y)\dot A_i(\vec x+\vec y)+K_2(\vec y)\e_{ijk}\pa_j
    \dot A_k(\vec x+\vec y)\Big)\ ,}\eqno(5.6)$$
where $K_i(\vec y)$ and $L_i(\vec y)$ now obey the following
symmetry conditions:
    $$K_i(-\vec y)=K_i(\vec y)\quad,\qquad L_i(-\vec
    y)=-L_i(\vec y)\ .\eqno(5.7)$$
Thus, we indeed obtained a transformation rule that leaves the
theory manifestly rotationally covariant.

Lorentz invariance is far less trivial to establish for this
system. In general, transversality of the fields is guaranteed
only if a Lorentz transformation is associated with a Weyl gauge
transformation. Remember that the transformation (5.6) is defined
in a given Cauchy surface $t=0$. So, let us start with a given
field $A_i(\vec x,t)$ with $\pa_i A_i=0$ and $A_0=0$. After an
infinitesimal Lorentz transformation in the $z$-direction, we have
$A_\m\ra A'_\m=A_\m+\e\d A_\m$, with $\e$ infinitesimal, and
    $$\eqalign{\d A_{1,2}&=z\dot A_{1,2}+t\,\pa_3 A_{1,2}+\pa_{1,2}\L\ ,\cr
    \d A_3&=z\dot A_3+ t\,\pa_3 A_3+\pa_3\L\ ,\cr
    \d A_0&=A_3+\pa_0\L\ .}\eqno(5.8)$$

We find
    $$\pa_i\d A_i=\dot A_3+\D\L\,,\eqno(5.9)$$
so, since we want the transformed fields to remain transverse, we
must choose the gauge transformation $\L$ to be
    $$\L=-\D^{-1}\dot A_3\,,\eqno(5.10)$$ which also implies
    $$\pa_0\L=-\D^{-1}\pa_0^2A_3=-A_3\,,\eqno(5.11)$$
so that $A_0$ also vanishes after the transformation. Thus we
find, at $t=0$,
    $$\d A_i=z\dot A_i-\D^{-1}\pa_i\dot A_3\quad,\qquad \d\dot
    A_i=z\D A_i+\pa_3 A_i-\pa_i A_3\ .\eqno(5.12)$$

Let us write our gauge transformation (5.6) in the following
shortcut notation,
    $$\eqalign{\tl A_i&=K_1A_i+K_2\,\e_{ijk}\,\pa_jA_k+L_1\dot
    A_i+L_2\,\e_{ijk}\,\pa_j\dot A_k\ ,\cr
    \tl{\dot A_i}&=L_1\D A_i+L_2\,\e_{ijk}\,\pa_j\D A_k+K_1\dot
    A_i+K_2\,\e_{ijk}\,\pa_j\dot A_k\ ,}\eqno(5.13)$$
After the Lorentz transformation (5.12), we have
    $$\eqalign{\tl A'_i=\tl A_i+\e\d\tl A_i\quad,&\qquad
    K'_i=K_i+\e\d K_i\ ,\cr
    \tl{\dot A}'_i=\tl{\dot A}_i+\e\d\tl{\dot A}_i\quad,&
    \qquad L'_i=L_i+\e\d L_i\ ,}\eqno(5.14)$$ where the $\d\tl
A_i$ and $\d\tl{\dot A_i}$ obey equations similar to (5.12). After
a little algebra, where we have to use the fact that the fields
are transverse,
    $$\pa_i A_i=\pa_i\dot A_i=0\,,\eqno(5.15)$$
we find that both of the equations (5.13) are obeyed by the
Lorentz transformed fields iff
    $$\eqalign{\d K_1=(z\,L_1-L_1\,z)\D-L_1\,\pa_3\,\quad,&\qquad
    \d L_1=z\,K_1-K_1\,z\ ,\cr
    \d K_2=(z\,L_2-L_2\,z)\D-2L_2\,\pa_3\quad,&\qquad \d
    L_2=z\,K_2-K_2\,z-K_2\,\D^{-1}\pa_3\ .}\eqno(5.16)$$

Eq.~(5.16) is written in operator notation. The kernel functions
$K_i(\vec y)$ and $L_i(\vec y)$ transform by
    $$\eqalign{\d K_1(\vec y)=-y_3\D L_1(\vec y)-\pa_3L_1(\vec y)\,\quad,&\qquad
    \d L_1(\vec y)=-y_3K_1(\vec y)\ ,\cr
    \d K_2(\vec y)=-y_3\D L_2(\vec y)-2\pa_3L_2(\vec y)\quad,&\qquad
    \d L_2(\vec y)=-y_3K_2(\vec y)-\pa_3\D^{-1}K_2(\vec y)\ .}\eqno(5.17)$$
It is important to note that the transformed kernels still obey
the (anti-)symmetry requirements (5.7). This proves that the
theory is indeed Lorentz covariant.

\newsect{6. Massless spinors.}
    Spinor fields cannot be treated in exactly the same manner. At
each Fourier mode of a Dirac field there are only a few quantum
degrees of freedom that can take only two values. At time steps
separated at a fixed distance $1/\pi\w$, where $\w=(\vec
k^2+\m^2)^{1/2}$, we may have elements of an ontological basis,
but not in between, so, unlike the bosonic case, we cannot take
the limit of continuous time.

However, an other approach was described in Ref\ref5. There, we
note that the first-quantized chiral wave equation can be given an
ontological basis. The massless, two-component case is described
by the Hamiltonian
    $$H=\vec\s\cdot\vec p\ .\eqno(6.1)$$

Take the following set of operators,
    $$\{\hat p,\quad \hat p\cdot\vec\s,\quad \hat p\cdot\vec x\}\ ,\eqno(6.2)$$
where $\hat p$ stands for the momentum {\it modulo\/} its length
and {\it modulo\/} its sign:
    $$\hat p=\pm \vec p/|p|\quad ;\qquad \hat p_x>0\ .\eqno(6.3)$$
These operators all commute with one another. Only the commutator
$[\,\hat p\cdot\vec x,\ \hat p\,]$ requires some scrutiny. In
momentum space we see
    $$[\,\vec p\cdot\vec x,\ \hat p\,]=i\left(\vec p\cdot{\pa\over\pa\vec
    p}\right)\,\hat p=0\,,\eqno(6.4)$$
because $\hat p$ has unit length; its length does not change under
dilatations. Not only does the set (6.2) commute with one another
at fixed time $t$, the operators commute at all times. This is
because $\hat p$ and $\hat p\cdot\vec\s$ commute with the
Hamiltonian, whereas
    $$\hat p\cdot\vec x(t)=\hat p\cdot\vec x(0)+\hat
    p\cdot\s\,t\,.\eqno(6.5)$$

For this reason, the set of basis elements in which the set of
operators (6.2) are diagonal, evolve by permutation, and we can
say that the evolution of these elements is deterministic.

If the particles do not interact, the numbers of particles and
antiparticles are fixed in time, and therefore also the second
quantized theory is deterministic. The only caveat here, is the
fact that the negative energy states must (nearly) all be
occupied; they form the Dirac sea. Since energy is not an
ontological observable here, the process of filling the Dirac sea
is a delicate one, and therefore it is advised to introduce a
cut-off: one may assume that the values of $\hat p\cdot\vec x$
form a dense but discrete lattice. The time required to hop from
one lattice point to the next is then a (tiny) time quantum. If
furthermore, at every value of the unit vector $\hat p$, we
introduce an infrared cut-off as well then the number of states in
the Dirac sea is finite, and filling the Dirac sea is
straightforward.

As in the bosonic case, discreteness in time leads to an ambiguity
in the Hamiltonian due to periodicity. In this case the most
obvious choice of the Hamiltonian is the symmetric one, half of
the first-quantized states having positive energy and half
negative.

We see that there is some similarity with the bosonic case. For
fermions, we may consider $\hat p\cdot\vec x$ to be one of the
three coordinates of a `classical particle', that carries with it
the value of $\hat p$ as a flag. Replacing the particle by
substituting its coordinates $\vec x$ by $\vec x+\vec y$ with
$\hat p\cdot\vec y=0$ ({\it i.e.\/}, a transverse displacement),
may here also be seen as a gauge transformation. All observables
must be invariant under this gauge transformation.

\newsect{7. Dissipation.}

The classical counterparts of the quantized models discussed here
are time-reversible, just as the quantum theories themselves. If
we restrict ourselves to gauge-invariant observables, the
classical theories described here are mathematically equivalent to
the quantum systems, the latter being familiar but very simple
quantum field models.

However, one might be inclined to go one step further. The large
gauge groups may in practice be difficult to incorporate in more
complicated ---~hence more interesting~--- settings, ones where
interactions and symmetry patterns are more realistic. This is why
it may be of importance to find an {\it interpretation\/} of our
gauge invariance. Consider the set of infinitesimal gauge
transformations out of which the finite ones can be generated. It
could be that these represent motions that cannot be predicted or
tracked back into the past: information concerning these motions
is lost\ref{5, 6}. Our fermionic particles, for instance, perform
random, uncontrollable steps sideways, while the projection of
their motion along the direction of $\hat p$ varies uniformly in
time.

Viewed from this perspective, it seems that we end up with
perfectly reasonable hidden variable theories. Those properties of
the system that can be followed during macroscopic time intervals
form {\it equivalence classes\/} of states.

In the models presented here, the equivalence classes are large,
but they typically multiply the number of degrees of freedom with
a factor two or so. In quantum gravity, we expect equivalence
classes much larger than that. In the vicinity of black holes,
only the degrees of freedom at the horizon represent observables;
all other degrees of freedom appear not to be independent of
those. For the interpretation and understanding of black holes,
this has been a formidable obstacle. If, however, the basis of our
physical Hilbert space can be seen to be spanned by equivalence
classes, then we can simply infer that black holes may also be
represented as classical objects, where equivalence classes are
large because information is divulged by the horizon.

\newsect{References.}

\item{1.} G. 't Hooft,  ``On the quantization of space and time",
in Proc. of the 4th seminar on Quantum Gravity, May 25-29, 1987,
Moscow, USSR. Eds. M.A. Markov, World Scientific, Singapore, New
Jersey, Hong Kong, 1988, pp. 551-567, and in: {\it Themes in
Contemporary Physics II. Essays in honor of Julian Schwinger's
70th birthday}. Eds. S.~Deser and R.J.~Finkenstein. World
Scientific, Singapore (1989) 77-89. \br See also: G.~'t~Hooft,
``Dimensional reduction in quantum gravity". In {\it
Salamfestschrift: a collection of talks}, World Scientific Series
in 20th Century Physics, vol. 4, ed. A. Ali, J. Ellis and S.
Randjbar-Daemi (World Scientific, 1993), gr-qc/9310026.

\item{2.} L.~Susskind, {\it J.~Math.~Phys. \bf 36} (1995) 6377,
hep-th/9409089.

\item{3.} G.~'t~Hooft, ``Determinism in Free Bosons", SPIN-2001/07,
hep-th/0104080.

\item{4.} G.~'t~Hooft, ``How Does God Play Dice? (Pre-)Determinism at the Planck
Scale", an Essay in honour of John S. Bell, SPIN-2001/09 /
ITP-UU-01/15, hep-th/0104219.

\item{5.} G.~'t~Hooft, ``Quantum Gravity as a Dissipative Deterministic System",
SPIN-1999/07, gr-qc/9903084; {\it Class.~Quant.~Grav. \bf 16}
(1999) 3263; ``Determinism and Dissipation in Quantum Gravity",
presented at {\it Basics and Highlights in Fundamental Physics},
Erice, August 1999, SPIN-2000/07, hep-th/0003005.

\item{6.} M.~Blasone, P.~Jizba and G.~Vitiello, ``Dissipation and
Quantization", hep-th/0007138.

 \bye